\documentclass[aps,prl,twocolumn,superscriptaddress,showpacs,showkeys]{revtex4-1}
\usepackage{graphicx}
\usepackage[abs]{overpic}
\usepackage{subfigure}
\newcommand{\bra}[1] {\langle #1 |}
\newcommand{\ket}[1] {| #1 \rangle}

\begin{document}

\title{The temperature shifts of the resonances of the NV$^-$ center in diamond}
\author{Marcus W. Doherty}
\email{marcus.doherty@anu.edu.au}
\affiliation{Laser Physics Centre, Research School of Physics and Engineering, Australian National University, Australian Capital Territory 0200, Australia.}
\author{Victor M. Acosta}
\altaffiliation[Present address: ]{Google [x], 1600 Ampitheatre Pkwy, Mountain View, CA 94043, USA.}
\affiliation{Department of Physics, University of California, Berkeley, CA 94720-7300, USA.}
\author{Andrey Jarmola}
\affiliation{Department of Physics, University of California, Berkeley, CA 94720-7300, USA.}
\author{Michael S.J. Barson}
\affiliation{Laser Physics Centre, Research School of Physics and Engineering, Australian National University, Australian Capital Territory 0200, Australia.}
\author{Neil B. Manson}
\affiliation{Laser Physics Centre, Research School of Physics and Engineering, Australian National University, Australian Capital Territory 0200, Australia.}
\author{Dmitry Budker}
\affiliation{Department of Physics, University of California, Berkeley, CA 94720-7300, USA.}
\affiliation{Nuclear Science Division, Lawrence Berkeley Laboratory, Berkeley CA 94720, USA.}
\author{Lloyd C.L. Hollenberg}
\affiliation{School of Physics, University of Melbourne, Victoria 3010, Australia.}

\date{\today}

\begin{abstract}
Significant attention has been recently focussed on the realization of high precision nano-thermometry using the spin-resonance temperature shift of the negatively charged nitrogen-vacancy (NV$^-$) center in diamond. However, the precise physical origins of the temperature shift is yet to be understood. Here, the shifts of the center's optical and spin resonances are observed and a model is developed that identifies the origin of each shift to be a combination of thermal expansion and electron-phonon interactions. Our results provide new insight into the center's vibronic properties and reveal implications for NV$^-$ thermometry.
\end{abstract}

\pacs{63.20.kp, 61.72.jn, 76.70.hb}

\maketitle

The negatively charged nitrogen-vacancy (NV$^-$) center in diamond \cite{review} is an important quantum technology platform for a range of new applications exploiting quantum coherence. Beyond quantum information processing, the prospect of employing the NV$^-$ center as a room temperature nanoscale electric and magnetic field sensor has attracted considerable interest \cite{efield,mag1,mag2,mag3,mag4,biomag}. Recently, the effects of temperature on the center's ground state spin resonance have been investigated \cite{acostashift1}, which enabled the influence of temperature on existing NV$^-$ metrology applications to be characterized and new thermometry applications to be proposed \cite{toyli12,acostashift1,acostashift2,chenshift} and demonstrated \cite{toyli13,neumann13,kucsko13}. However, the temperature shift of the center's spin resonance is not well understood and previous attempts at modelling the shift have been largely unsuccessful \cite{acostashift1,acostashift2,chenshift}. It is evident that the implementation of the NV$^-$ center as a nano-thermometer, magnetometer or electrometer requires a thorough understanding of the temperature shifts of its resonances, particularly if these implementations are designed for ambient conditions \cite{fang13}. Here, the temperature shifts of the center's visible, infrared and spin resonances are observed and a model is developed that identifies the origin of each shift to be a combination of thermal expansion and electron-phonon interactions. This new insight reveals implications for NV$^-$ metrology.

\begin{figure}[hbtp]
\begin{center}
\mbox{
\subfigure[]{
\includegraphics[width=0.4\columnwidth] {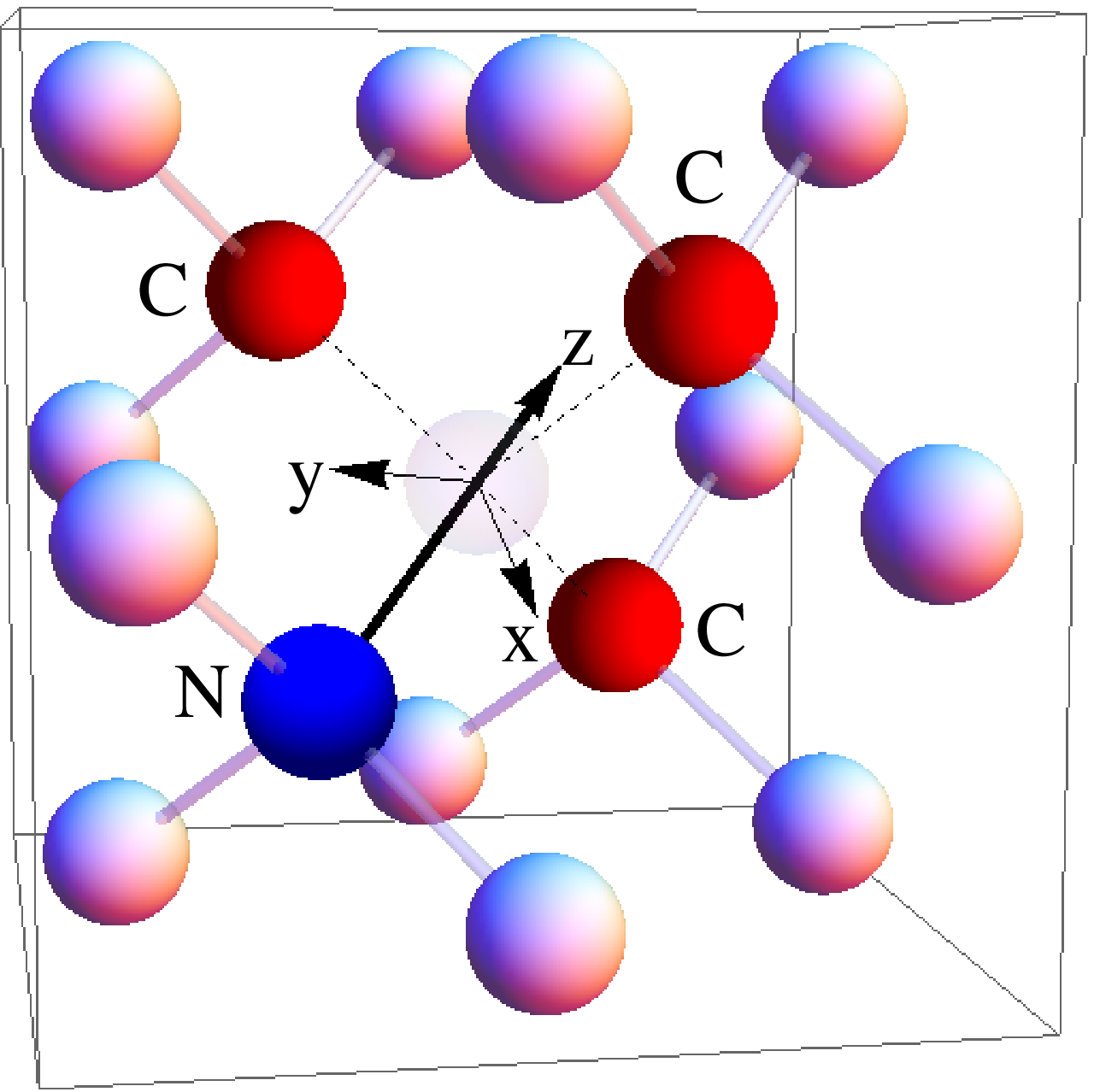}}
\subfigure[]{\includegraphics[width=0.45\columnwidth] {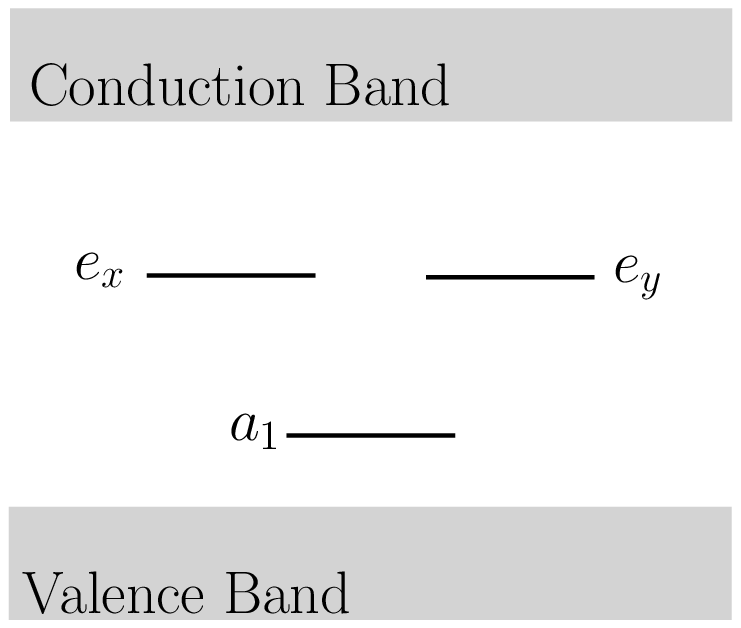}}
}
\mbox{
\subfigure[]{\includegraphics[width=0.75\columnwidth] {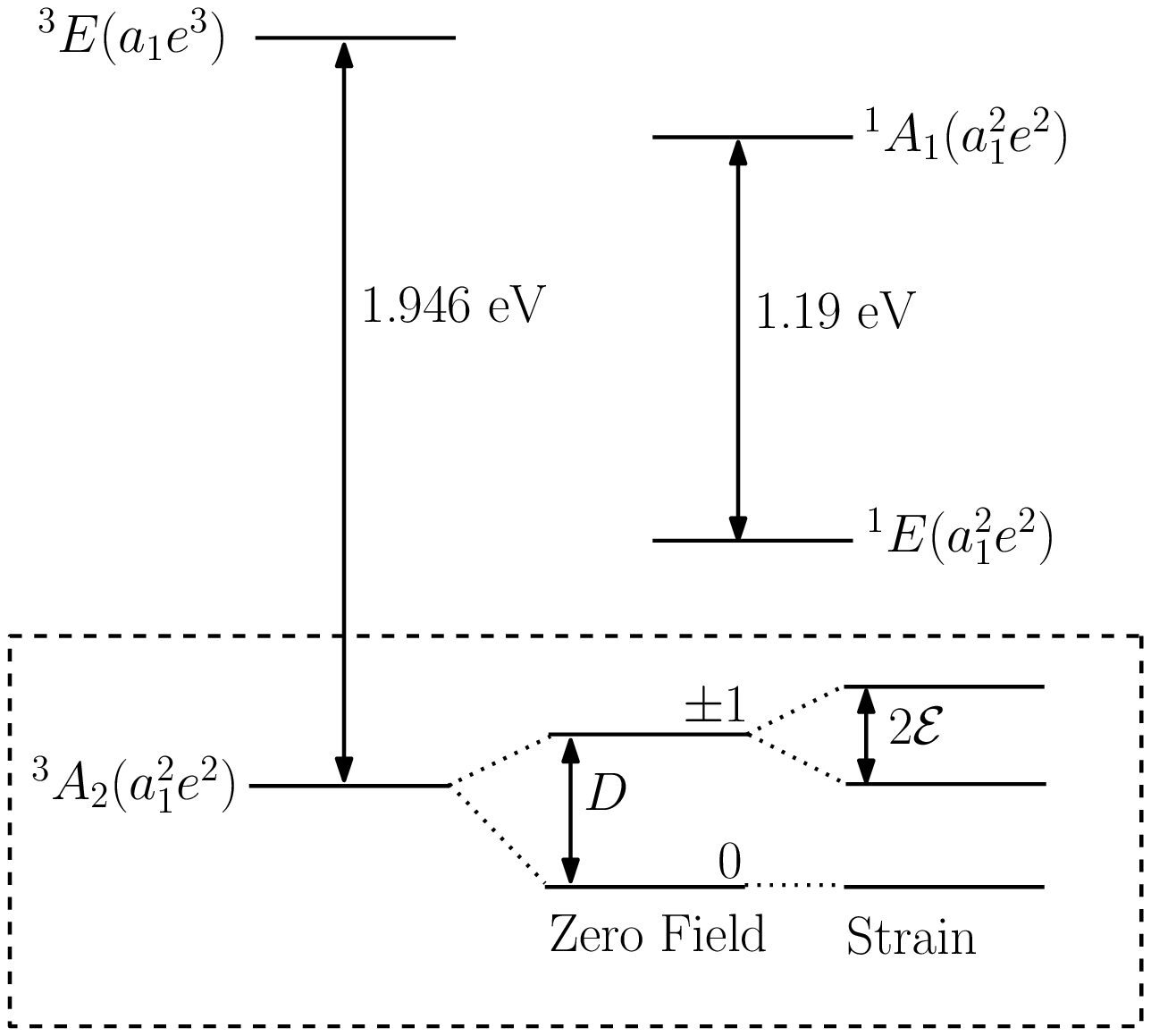}}
}
\caption{(color online) (a) Schematic of the NV center depicting the vacancy, the nearest neighbor carbon atoms, the substitutional nitrogen atom, the next-to-nearest carbon neighbors and the adopted coordinate system (z axis aligned with the $C_{3v}$ axis of the center and the $x$ axis is contained in one of the center's mirror planes). (b) The NV$^-$ one-electron orbital level structure depicting the diamond valence and conduction bands and the three defect orbitals ($a_1$, $e_x$, and $e_y$) within the bandgap. (c) Schematic of the center's many-electron electronic structure, including the low-temperature visible $E_\mathrm{V}\sim$1.946 eV and infrared $E_\mathrm{IR}\sim$1.19 eV ZPL energies. The electronic configurations of the many-electron levels are indicated in parentheses. Inset: The fine structure of the ground $^3A_2$ level: at zero field with a single splitting $D\sim2.88$ GHz (low temperature); and under symmetry lowering strain, with an additional strain dependent splitting $2{\cal E}$. }
\label{fig:fig1}
\end{center}
\end{figure}

The NV$^-$ center is a $C_{3v}$ point defect in diamond consisting of a substitutional nitrogen atom adjacent to a carbon vacancy that has trapped an additional electron (refer to Fig. \ref{fig:fig1}a). As depicted in Fig. \ref{fig:fig1}b, the one-electron orbital level structure of the NV$^-$ center contains three defect orbital levels ($a_1$, $e_x$ and $e_y$) deep within the diamond bandgap. Electron paramagnetic resonance (EPR) observations and ab initio calculations indicate that these defect orbitals are highly localized to the center \cite{he93,felton09,larrson08,gali08,hossain08}. Figure \ref{fig:fig1}c shows the center's many-electron electronic structure generated by the occupation of the three defect orbitals by four electrons \cite{doherty11,maze11}, including the low-temperature zero phonon line (ZPL) energies of the visible ($E_\mathrm{V}\sim$1.946 eV)  \cite{davies76} and infrared ($E_\mathrm{IR}\sim$1.19 eV) \cite{rogers08,acosta10b,manson10} transitions. The energy separations of the spin triplet and singlet levels (${^3}A_2\leftrightarrow{^1}E$ and ${^1}A_1\leftrightarrow{^3}E$) are unknown.

As depicted in the inset of Fig. \ref{fig:fig1}c, the ground $^3A_2$ level exhibits a zero-field fine structure splitting between the $m_s=0$ and $\pm1$ spin sub-levels of $D\sim2.88$ GHz (low temperature) due principally to electron spin-spin interaction \cite{loubser78}.  Under crystal strain that distorts the $C_{3v}$ symmetry of the center, the $m_s=\pm1$ sub-levels are mixed and their degeneracy is lifted in the absence of a magnetic field. The spin-Hamiltonian that describes the $^3A_2$ fine structure is
\begin{equation}
H = D[S_z^2-S(S+1)/3]+{\cal E}(S_x^2-S_y^2),
\end{equation}
where ${\cal E}$ is the strain parameter, the $S=1$ spin operators are dimensionless and the $z$ coordinate axis coincides with the center's $C_{3v}$ symmetry axis (see Fig. \ref{fig:fig1}a).

The spin of the ground $^3A_2$ level is optically polarized due to spin-selective non-radiative intersystem crossings (ISC) that preferentially depopulate the $m_s=\pm1$ sub-levels and populate the $m_s=0$ sub-level \cite{review}. The ISC also lead to spin-dependent optical fluorescence that enable the ground state spin to be measured and the performance of optically detected magnetic resonance (ODMR) \cite{review}. The non-radiative ISC of the NV$^-$ center are not fully understood, but they are currently believed to be the combined result of spin-orbit coupling of the lowest energy triplet ($^3A_2$, $^3E$) and singlet ($^1A_1$, $^1E$) levels and electron-phonon interactions \cite{review}.

There were extensive temperature studies of the NV$^-$ center conducted at the time of its first identification \cite{davies74,collins83}. In his early vibronic study \cite{davies74}, Davies observed the temperature shifts of the visible ZPL energy accompanied by changes in its homogeneous linewidth and luminescence intensity. Davies provided an accurate vibronic model of these temperature effects by considering both the response of the ZPL to the stress induced by thermal expansion of the bulk lattice as well as electron-phonon interactions at the center. The temperature dependence of the ground state spin resonance was studied by Acosta et al \cite{acostashift1,acostashift2} who observed a shift in the $D$ parameter. Acosta et al demonstrated that the shift of $D$ was not consistent with a model based purely on the $\sim\langle1/r^3\rangle$ spatial dependence of electron spin-spin interaction and the thermal expansion of the bulk crystal lattice. Chen et al \cite{chenshift} later observed that the temperature dependence of the shifts of $D$ and the visible ZPL energy were similar and proposed that the shifts were both due to some local thermal expansion that acted similarly on the center's visible and magnetic transitions. However, Chen et al did not add detail to their hypothesis, thereby leaving the explanation of the temperature shift of $D$ an unresolved problem. To our knowledge, the temperature shift of the infrared ZPL energy has not been reported previously.

Davies' model of the temperature shift of the visible ZPL is equally valid for the shifts of the infrared ZPL and $D$. The later is evident once it is recognised that the spin transitions of the ground state spin resonance are zero-phonon transitions (i.e. the observed spin resonance is a ZPL). The quasi-harmonic approximation is adopted in Davies' model so that the explicit treatment of vibrational anharmonicity is avoided (see supplementary material for further discussion). Instead, the anharmonic displacement of the mean positions of nuclei at higher vibrational quantum numbers is approximated by the static strain of thermal expansion. The anharmonic reduction of vibrational energy level spacings at higher quantum numbers is ignored, such that the vibrational modes remain harmonic and their frequencies are temperature independent.

\begin{figure}[hbtp]
\begin{center}
\includegraphics[width=1.0\columnwidth] {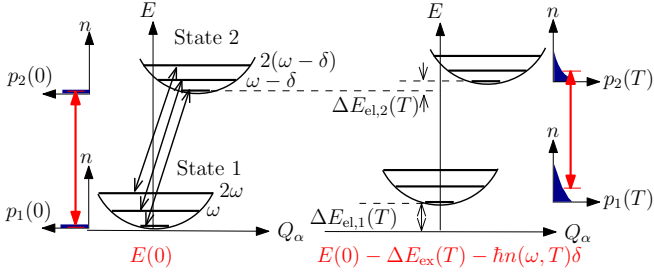}
\caption{(color online) Configuration coordinate diagrams depicting transitions between the vibronic manifolds of two electronic states in the limit of low temperature (left) and at higher temperature $T$ (right). $Q_\alpha$ is the mass-weighted nuclear displacement coordinate of the $\alpha^{th}$ vibrational mode with frequency $\omega$ in electronic state 1 and frequency $\omega-\delta$ in state 2. Harmonic vibrational potential-energy curves and corresponding vibrational levels for the electronic states are black-solid. $\Delta E_\mathrm{el,i}(T)$ is the change in the electronic energy of the $i^{th}$ level due to the thermal expansion. Example zero-phonon transitions are depicted by black solid arrows. $n$ is the vibrational quantum number and $p_i(n)$ is the thermal distribution of vibrational occupations of the $i^{th}$ electronic state, with the mean occupation denoted by a red horizontal line.  The thermal average zero-phonon transition energy (solid arrows) are depicted in red. }
\label{fig:configcoord}
\end{center}
\end{figure}

In Davies' model, there are two contributions to temperature shifts of ZPLs: (1) the electronic energies are perturbed by the strain of thermal expansion, and (2) the vibrational frequencies associated with different electronic levels differ. The later is a consequence of the vibrational potential-energy functions of the electronic levels having different curvatures. Equivalently, the differences in vibrational frequencies may be described as outcomes of quadratic electron-phonon interactions \cite{davies74}. Davies' model may be further understood by considering configuration-coordinate diagrams that depict transitions between the vibronic manifolds of two electronic states (see Fig. \ref{fig:configcoord}). Zero-phonon transitions occur between vibronic levels of the electronic states that do not differ in their vibrational quantum number. If the initial electronic state is in thermal equilibrium, the observed central ZPL energy is the thermal average of all zero-phonon transition energies.

As temperature increases, the equilibrium positions of the nuclei expand and the electronic energies shift, resulting in a common shift of all of the zero-phonon transition energies, and thus of the ZPL energy (see Fig. \ref{fig:configcoord}). Introducing the mass-weighted nuclear displacement coordinate $Q_\mathrm{ex}(T)$ of thermal expansion, the relative temperature shift of the electronic energies is
\begin{eqnarray}
\Delta E_\mathrm{ex}(T) = \left.\frac{\partial \Delta E_\mathrm{el}}{\partial Q_\mathrm{ex}}\right|_0Q_\mathrm{ex}(T),
\label{eq:expshift}
\end{eqnarray}
where the derivative is evaluated at $Q_\mathrm{ex}=0$ and $\Delta E_\mathrm{el}=E_\mathrm{el,1}-E_\mathrm{el,2}$ is the difference in electronic energies and $E_\mathrm{el,i}$ is the electronic energy of the $i^{th}$ electronic state. Since thermal expansion and hydrostatic pressure are intimately related, the contribution of thermal expansion to the ZPL shift is more simply \cite{davies74}
\begin{eqnarray}
\Delta E_\mathrm{ex}(T) = A P(T),
\end{eqnarray}
where $A$ is the hydrostatic pressure shift of the ZPL,  $P(T) = B\int_0^T e(t)dt$ is the pressure of thermal expansion, $B = 442$ GPa is the bulk modulus of diamond and $e(T)$ is the diamond volume expansion coefficient.

If the vibrational frequencies of the two electronic states differ, the energy of a zero-phonon transition will depend linearly on vibrational quantum number. Introducing the vibrational density of modes $\rho(\omega)$ and performing the thermal average, the contribution of electron-phonon interactions to the ZPL temperature shift is
\begin{eqnarray}
\Delta E_\mathrm{e-p} (T) = \hbar\int_0^\Omega n(\omega,T)\delta(\omega)\rho(\omega)d\omega,
\end{eqnarray}
where $n(\omega,T) = (e^{\hbar\omega/k_B T}-1)^{-1}$ is the Bose-Einstein distribution of vibrational occupations, $\Omega\sim165$ meV is the highest vibrational frequency (approximately the diamond Debye frequency), and $\delta(\omega)$ is the average vibrational frequency difference between the electronic states, which is defined by
\begin{eqnarray}
&&\delta(\omega)\rho(\omega)\nonumber \\
 && =  \sum_{\alpha:\omega_{1,\alpha}=\omega}(\omega_{1,\alpha}-\omega_{2,\alpha})\approx\frac{1}{2\omega}\sum_{\alpha:\omega_{1,\alpha}=\omega}\left.\frac{\partial^2 \Delta E_\mathrm{el}}{\partial Q_\alpha^2}\right|_0.
\label{eq:epshift}
\end{eqnarray}
In the above, $Q_\alpha$ and $\omega_{i,\alpha}$ are the  mass-weighted displacement coordinate and frequency of the $\alpha^{th}$ mode in the $i^{th}$ electronic state, respectively, the derivative is evaluated at $Q_\alpha=0$ and the sum is over all modes with frequency $\omega_{1,\alpha}=\omega$ \cite{supmat}. The total temperature shift of the ZPL is finally $\Delta E(T) = -\Delta E_\mathrm{ex}(T) - \Delta E_\mathrm{e-p} (T)$.

The hydrostatic pressure shifts $A$ of the visible, infrared and spin resonances of NV$^-$ have each been measured previously (see table \ref{tab:parameters}) \cite{review,davies74,doherty13}. So has the volume thermal expansion coefficient of diamond, which can be expressed as a power series $e(T)=\sum_{i=1}^4 e_i T^i$ terminating at $T^4$ for $T<300$ K \cite{texpansion}. Importantly, the thermal expansion coefficient depends on the purity of the diamond \cite{texpansion}, which is discussed further below. Consequently, $\Delta E_\mathrm{ex}(T)$ of each resonance can be predicted from previous experimental results.  Given $\Delta E_\mathrm{ex}(T)$, $\Delta E_\mathrm{e-p}(T)$ may be approximately determined for each resonance by first expanding $\delta(\omega)\rho(\omega)\approx \sum_{i=3}a_i\omega^i$, which via the evaluation of the integral in $\Delta E_\mathrm{e-p}(T)$, yields the temperature expansion $\Delta E_\mathrm{e-p}(T)\approx \sum_{i=4} b_i T^i$ \cite{supmat}. The expansion of $\delta(\omega)\rho(\omega)$ commences at $\omega^3$ because in the limit of $\omega\rightarrow0$, $\rho(\omega)\propto\omega^2$ (i.e. the Debye density) and for quadratic electron-phonon interactions $\delta(\omega)\propto\omega$ \cite{davies74,maradudin66}. Terminating the expansion of $\Delta E_\mathrm{e-p}(T)$ at $T^5$, the approximate expression for a temperature shift is
\begin{eqnarray}
\Delta E(T) &\approx& -\frac{e_1}{2}ABT^2-\frac{e_2}{3}ABT^3-\left(b_4+\frac{e_3}{4}AB\right)T^4 \nonumber \\
&& -\left(b_5+\frac{e_4}{5}AB\right)T^5.
\label{eq:expression}
\end{eqnarray}

\begin{table*}
\caption{\label{tab:parameters} Parameters of expression (\ref{eq:expression}) for the temperature shifts of the NV$^-$ visible, infrared and spin resonances. Only the quadratic electron-phonon interaction parameters ($b_4$, $b_5$) were free parameters in the least-squares fits depicted in Fig. \ref{fig:fits}. The thermal expansion parameters $e_i$ are derived from those obtained in Ref. \onlinecite{texpansion}.}
\begin{ruledtabular}
\begin{tabular}{llllllll}
Shift & $A$ & $\frac{ABe_1}{2}$ & $\frac{ABe_2}{3}$ & $\frac{ABe_3}{4}$ & $\frac{ABe_5}{5}$ & $b_4$ & $b_5$  \\
(Unit) & $\frac{\mathrm{Unit}}{\mathrm{GPa}}$ & $\mathrm{Unit}/\mathrm{T}^{2}$ & $\mathrm{Unit}/\mathrm{T}^{3}$ & $\mathrm{Unit}/\mathrm{T}^{4}$ & $\mathrm{Unit}/\mathrm{T}^{5}$ & $\mathrm{Unit}/\mathrm{T}^{4}$ & $\mathrm{Unit}/\mathrm{T}^{5}$   \\
\hline
$\Delta D$ (MHz) & 14.6\footnote{Reference \onlinecite{doherty13}.} & 39.7$\times10^{-7}$ & -91.6$\times10^{-9}$ & 70.6$\times10^{-11}$ & -60.0$\times10^{-14}$ & -18.7(4)$\times10^{-10}$ & 41(2)$\times10^{-13}$  \\
$\Delta E_O$ (meV) & 5.75$^\mathrm{a}$ & 15.6$\times10^{-7}$ & -36.1$\times10^{-9}$ & 27.9$\times10^{-11}$ & -23.7$\times10^{-14}$ & -8.0(8)$\times10^{-10}$ & 14(3)$\times10^{-13}$  \\
$\Delta E_{IR}$ (meV) & 1.45\footnote{Reference \onlinecite{review}.} & 3.95$\times10^{-7}$ & -9.12$\times10^{-9}$ & 7.03$\times10^{-11}$ & -5.97$\times10^{-14}$ & -1.9(7)$\times10^{-10}$ & 0.3(1)$\times10^{-13}$  \\
\end{tabular}
\end{ruledtabular}
\end{table*}

\begin{figure}[hbtp]
\begin{center}
\mbox{
\subfigure[]{
\includegraphics[width=0.5\columnwidth] {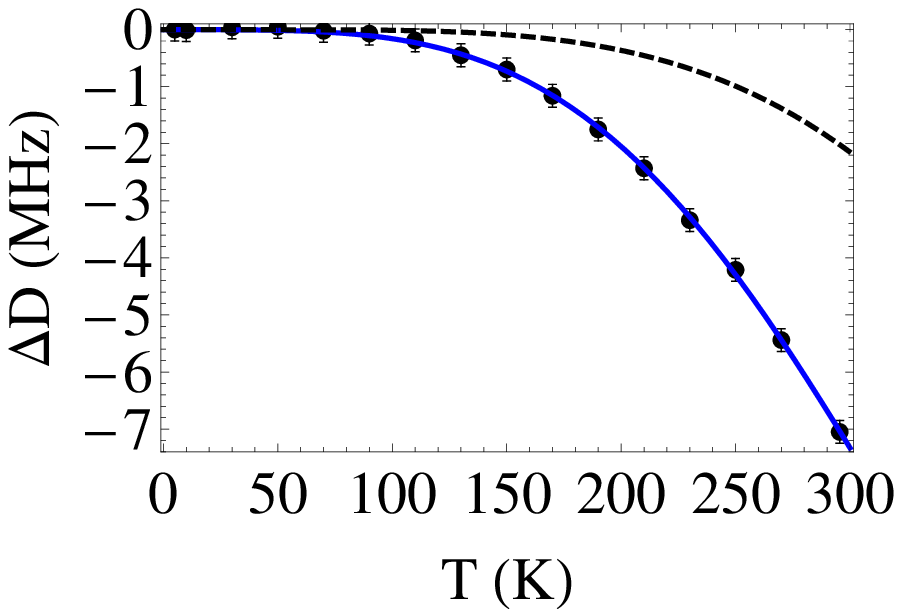}}
\subfigure[]{\includegraphics[width=0.5\columnwidth] {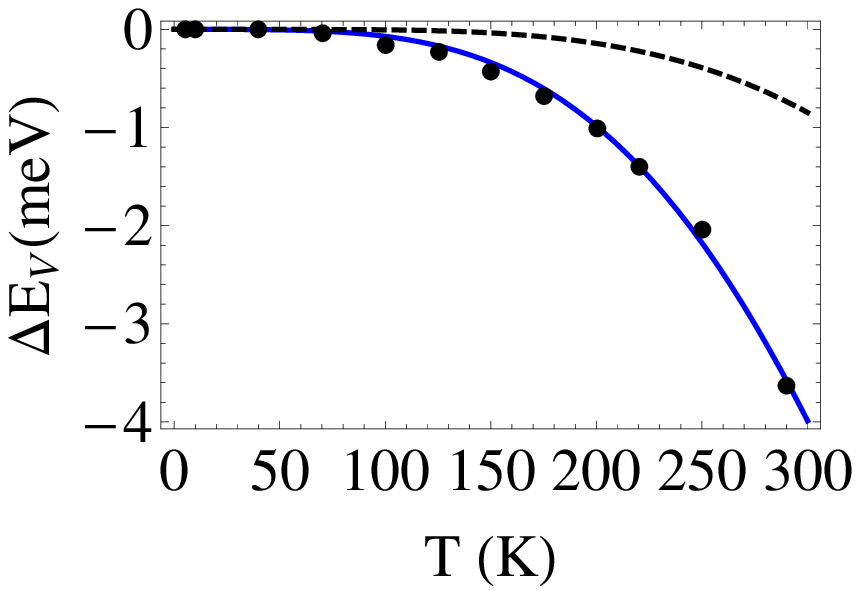}}
}
\mbox{
\subfigure[]{\includegraphics[width=0.5\columnwidth] {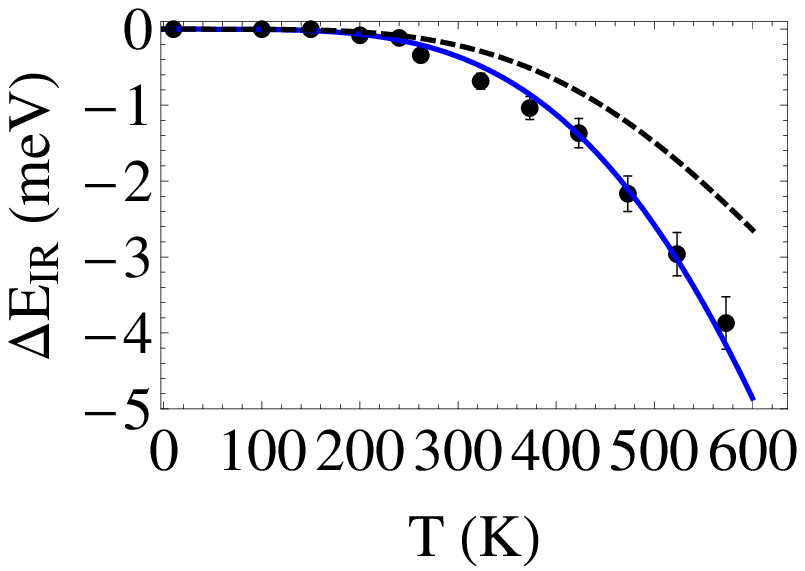}}
\subfigure[]{\includegraphics[width=0.5\columnwidth] {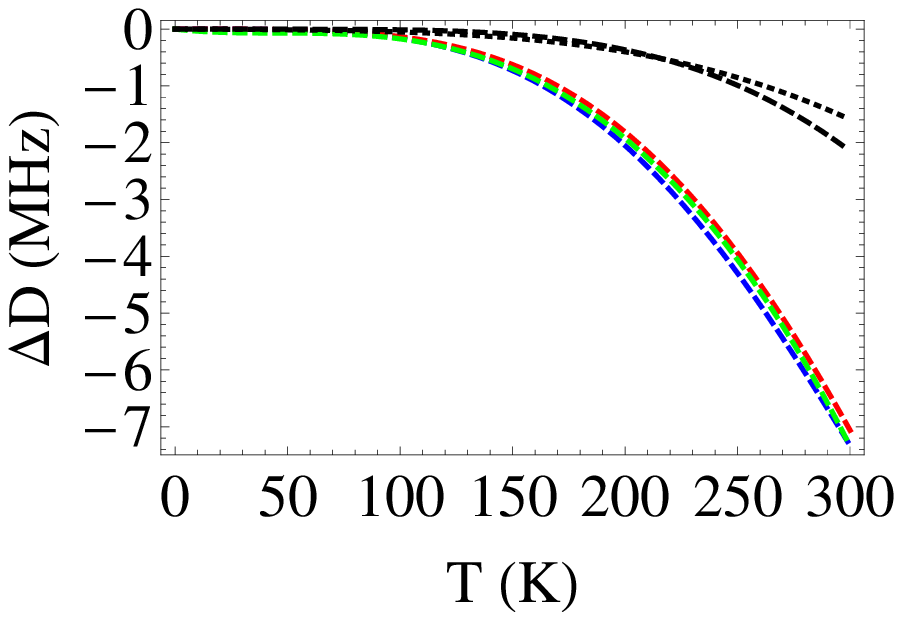}}
}
\caption{(color online) (a)-(c) The temperature shifts of the NV$^-$ spin, visible and infrared resonances, respectively (black points - measurements, blue curves - fit obtained using (\ref{eq:expression})). The contributions of thermal expansion $\Delta E_\mathrm{ex}(T)$ to each shift alone are depicted as dashed black curves. Experimental uncertainties are discussed in the supplementary material. (d) Comparison of the observed spin resonance shifts in bulk diamonds with different impurity nitrogen concentration [N]: green dashed - [N]$<$1 ppm (observed by Chen et al \cite{chenshift}), blue dashed - [N]$\sim0.4-4$ ppm, and red dashed - [N]$\sim40-60$ ppm. The thermal expansion contribution to the spin resonance shift calculated using the known thermal expansion coefficients of bulk diamonds with different impurity nitrogen concentration \cite{texpansion}: black dashed - very low concentration, black dotted - high concentration [N]$\sim$78 ppm.}
\label{fig:fits}
\end{center}
\end{figure}

In our experiments, we employed optical spectroscopy and ODMR techniques to measure the shifts of NV$^-$ ensemble resonances over the temperature range 5-300 K (see Ref. \onlinecite{supmat} for details). The shifts of the visible, infrared and spin resonances were measured in bulk high-pressure high-temperature (HPHT) type Ib samples containing similar nitrogen impurity (40-60 ppm). The shift of the spin resonance was also measured in another bulk HPHT sample with lower nitrogen impurity (0.4-4 ppm). Whilst the visible and infrared ZPLs were observed in emission, their respective emitting electronic levels are sufficiently long-lived for thermal equilibrium of their vibrational levels to be achieved within their lifetime \cite{review}. Figure \ref{fig:fits} depicts our observations together with fits using the expression (\ref{eq:expression}) and the shifts $\Delta E_\mathrm{ex}(T)$ predicted purely by thermal expansion (see table \ref{tab:parameters} for fit parameters). It is clear that at their respective powers of $T$, the parameters of $\Delta E_\mathrm{e-p}(T)$ are an order of magnitude larger than the parameters of $\Delta E_\mathrm{ex}(T)$, which demonstrates that the inclusion of electron-phonon interactions is necessary to explain each of the shifts.

The electronic model of the center can be applied to gain further physical insight (see supplementary material for details). Equations (\ref{eq:expshift}) and (\ref{eq:epshift}) demonstrate that the thermal expansion and electron-phonon contributions to the ZPL temperature shifts are determined by the first and second derivatives of the electronic energy differences $\Delta E_\mathrm{el}$ with respect to nuclear displacement. As the nuclei are displaced, the electron-nucleus electrostatic interactions are modified, resulting in changes of the center's orbitals and their energies. The changes to the orbitals in turn perturb electron-electron interactions. The $^3A_2(a_1^2e^2)\rightarrow$$^3E(a_1e^3)$ transitions of the visible ZPL involve a change in electronic configuration. Thus, the temperature shift of the visible ZPL depends on how defect orbital energies as well as electron-electron electostatic repulsion vary with nuclear displacement. The $^1E(a_1^2e^2)\rightarrow$$^1A_1(a_1^2e^2)$ transitions of the infrared ZPL and spin transitions of the $^3A_2(a_1^2e^2)$ ground state, however, do not involve a change in electronic configuration, and thus only depend on how electron-electron electrostatic repulsion and spin-spin interaction, respectively, vary with nuclear displacement due to changes in the defect orbitals. Given previous study of the visible ZPL \cite{davies74}, we will concentrate on the temperature shifts of the infrared ZPL and $D$.

Picturing the defect orbitals as linear combinations of atomic orbitals, nuclear displacement changes the defect orbitals in two ways: (1) the atomic orbitals are displaced, and (2) the linear combinations are modified in response to changed electrostatic interactions \cite{doherty13}. The molecular model of the NV$^-$ center \cite{doherty11,maze11,doherty12} yields the following expressions for the electronic energy differences $\Delta E_\mathrm{el}$ corresponding to the infrared and spin resonances
\begin{eqnarray}
\Delta E_\mathrm{el}^\mathrm{IR} & = & 2K \bra{e_x(\vec{r}_1)e_y(\vec{r}_2)}\frac{1}{r_{12}}\ket{e_y(\vec{r}_1)e_x(\vec{r}_2)} \approx \frac{K}{3}\eta^2\langle\frac{1}{r_{12}}\rangle \nonumber \\
D & = & C \bra{e_x(\vec{r}_1)e_y(\vec{r}_2)}\frac{1}{r_{12}^3}-\frac{3z_{12}^2}{r_{12}^5}\left[\ket{e_x(\vec{r}_1)e_y(\vec{r}_2)}\right. \nonumber \\
&& \left.-\ket{e_y(\vec{r}_1)e_x(\vec{r}_2)}\right] \approx C\eta^2\langle\frac{1}{r_{12}^3}-\frac{3z_{12}^2}{r_{12}^5}\rangle
\end{eqnarray}
where $K$ and $C$ are constants (see supplementary material for definition), $\vec{r}_i = x_i\hat{\vec{x}}+y_i\hat{\vec{y}}+z_i\hat{\vec{z}}$ is the position of the $i^{th}$ electron, $r_{12}=|\vec{r}_2-\vec{r}_1|$, $z_{12}=z_2-z_1$,  $\langle\ldots\rangle=\bra{c_1(\vec{r}_1)c_2(\vec{r}_2)}\ldots\ket{c_1(\vec{r}_1)c_2(\vec{r}_2)}$, $c_i$  are the dangling sp$^3$ atomic orbitals of the vacancy's three nearest neighbor carbon atoms, and $\eta=\sum_{i=1}^3|<c_i|e_x>|^2\sim0.84$ \cite{felton09}. The approximations in the above are the neglect of atomic orbitals other than the atomic orbitals of the three nearest neighbor carbon atoms and the neglect of orbital overlaps. The approximate expressions demonstrate that there are two factors: (1) the electron density $\eta$ associated with the three carbon atoms and (2) the expectation value of the interaction between the dangling sp$^3$ electron densities of two of the carbon atoms.  As defined in equations (\ref{eq:expshift}) and (\ref{eq:epshift}), the parameters of table \ref{tab:parameters} are directly related to the first and second derivatives of these factors with nuclear displacement. These parameters therefore provide insight into how the electron density and interaction expectation values depend on nuclear displacement. Future \textit{ab initio} calculations of these parameters will yield further insight into the center's vibronic properties.

The principal implication for NV$^-$ metrology that our model reveals concerns variation between nanodiamond NV$^-$ centers. Figure \ref{fig:fits}(d) compares the reported $\Delta D(T)$ and calculated $\Delta E_\mathrm{ex}(T)$ contribution of centers in bulk diamonds with different nitrogen impurity. Figure \ref{fig:fits}(d) demonstrates that observed $\Delta D(T)$ are similar, implying that due to the significant contribution of $\Delta E_\mathrm{e-p}(T)$, $\Delta D(T)$ will vary little between bulk diamonds with different nitrogen impurity. However, in sufficiently small nanodiamonds, where the density $\rho(\omega)$ of vibrational modes is modified, the contribution of $\Delta E_\mathrm{e-p}(T)$ may vary significantly. Furthermore, in such nanodiamonds, the thermal expansion coefficient is likely to also be significantly modified due to structural defects and surface morphology. If $\Delta D(T)$ varies significantly between each nanodiamond NV$^-$ thermometer, then either careful consistency in the fabrication of NV$^-$ thermometers or calibration and marking of each NV$^-$ thermometer is necessary for the successful implementation of NV$^-$ nanothermometry. Further investigations are clearly required.

\begin{acknowledgements}
This work was supported by the Australian Research Council under the Discovery Project scheme (DP0986635 and DP120102232), the NSF, the AFOSR/DARPA QuASAR program, NATO SFP, and IMOD. M.W.D. wishes to acknowledge the David Hay Memorial Fund.
\end{acknowledgements}

\end{document}